\documentclass[twoside,english]{iopart}
\usepackage{geometry}
\usepackage{textcomp}
\usepackage{amsbsy}
\usepackage{amstext}
\usepackage{graphicx}
\usepackage{esint}
\usepackage{color}

\makeatletter
\usepackage{iopams}
\usepackage{setstack}

\def\newblock{\hskip .11em plus .33em minus .07em}

\newcommand{\eqref}[1]{(\ref{#1})}

\makeatother

\usepackage{babel}
\begin{document}

\title[Giant fluctuations in a flocking epithelium]{Giant fluctuations and structural effects in a flocking epithelium}

\author{{\normalsize{}Fabio Giavazzi$^{1}$, Chiara Malinverno$^{2}$, Salvatore
Corallino$^{2}$, Francesco Ginelli$^{3}$, Giorgio Scita$^{4,2}$
and Roberto Cerbino$^{1}$}}

\address{1 Universit\'a degli Studi di Milano, Dipartimento di Biotecnologie
Mediche e Medicina Traslazionale, I-20090 Segrate, Italy}

\address{2 IFOM-FIRC Institute of Molecular Oncology, I-20139 Milano, Italy}

\address{3 SUPA, Institute for Complex Systems and Mathematical Biology and Department of Physics, University of Aberdeen, Aberdeen, AB24 3UE, United Kingdom}

\address{4 Universit\'a degli Studi di Milano, Dipartimento di Oncologia e Emato-Oncologia,
I-20133 Milano, Italy}

\ead{fabio.giavazzi@unimi.it}
\begin{abstract}
Epithelial cells cultured in a monolayer are very motile in isolation but reach a near-jammed state when mitotic division increases their number above a critical threshold. We have recently shown that a monolayer can be reawakened by over-expression of a single protein, RAB5A, a master regulator of endocytosis. This reawakening of motility was explained in term of a flocking transition that promotes the emergence of a large-scale collective migratory pattern. Here we focus on the impact of this reawakening on the structural properties of the monolayer. We find that the unjammed monolayer is characterised by a fluidisation at the single cell level and by enhanced non-equilibrium 
large-scale number fluctuations at a larger length scale. Also with the help of numerical simulations, we trace back the origin of these fluctuations to the 
self-propelled active nature of the constituents and to the existence of a local alignment mechanism, leading to the spontaneous breaking of the orientational symmetry.
\end{abstract}

%\submitto{\JPD }
\maketitle

\section{Introduction}

Fluctuations are of key importance in statistical physics, a paradigmatic example being the erratic motion displayed by a colloidal particle as a consequence of the random collisions with the molecules of the solvent. In systems at equilibrium, however, thermal fluctuations are usually small and — most importantly — obey a normal statistics, so that in many instances they can be easily accounted for or even safely neglected when the system is described at the macroscopic level. A familiar but remarkable exception is represented by thermodynamic systems close to a critical point \cite{criticalbook}, such as a binary mixture close to its consolution point. Due to the vanishingly small osmotic compressibility, concentration fluctuations become so large to be visible macroscopically, opening to the possibility of accessing important thermodynamic quantities with light scattering \cite{Berne:1976yu,Sengers2009} or quantitative microscopy \cite{Giavazzi:2014it,Giavazzi2016} methods. Far from equilibrium, fluctuations can become anomalously large in a generic region of the phase diagram, thus without the need of fine tuning the system close to a critical point. This can occur for instance in inert systems that are in a non-equilibrium state because of the presence of a macroscopic gradient \cite{DeZarate:2006qe,Giavazzi:2016fr} or also for \textit{active matter} systems, composed by individual entities intrinsically kept far from equilibrium by energy being injected and dissipated at the microscopic, single component level \cite{Rama2010, RevModPhys.85.1143}. A notable class of active matter is represented by living matter, typically fueled by internal
biochemical processes. Living active matter examples range from sub-cellular biofilaments displaced by motor proteins \cite{Schaller2010_,Schaller2011,Sumino_,sanchez_} to bacterial colonies \cite{Swinney2010, Peruani}, insects swarms \cite{mosquitos} and superior social vertebrates \cite{birds1,birds2,birds3,sheep}.

All these systems exhibit remarkable fluctuations in significant variables, the most paradigmatic example being represented by the so called giant number fluctuations (GNF), a hallmark of the inhomogeneous spatial distribution of the individual components in the system \cite{RevModPhys.85.1143}. A very direct way to quantify GNF is to consider the fluctuation $\delta N=N-\left\langle N\right\rangle $ in the number $N$ of moving entities contained in a region of prescribed size around its mean value $\left\langle N\right\rangle $. If the entities are randomly and independently distributed in space (the so-called spatial Poisson process) the hypotheses of the central limit theorem (CLT) are fulfilled and the mean squared amplitude of the number fluctuation is expected to be equal to $\left\langle \delta N^{2}\right\rangle \sim\left\langle N\right\rangle $. If, on the other hand, long-ranged positional correlations are present, significant deviations form this simple behaviour are expected and different scaling exponents are observed for $\left\langle \delta N^{2}\right\rangle$ as a function of $\left\langle N\right\rangle $ \cite{Gabrielli}. GNF have been shown to be a simple yet powerful and universal indicator of mesoscopic scale fluctuations in active matter systems \cite{RevModPhys.85.1143}, such as for instance active granular materials \cite{Desaigne2010}, bacteria \cite{Swinney2010} and cells \cite{Garcia,Zehnder}.
 
Truly asymptotic GNF, on the other hand, characterise active matter systems that exhibit long ranged order, such as polar
\textit{flocking}, a form of collective motility displayed, for instance, by birds flying in flocks and that can be understood through a minimal model 
(known as the Vicsek model) combining self-propulsion and effective alignment interactions \cite{Vicsek95}.
Hydrodynamic theory results \cite{TonerTu1,TonerTu2}, and numerical simulations based on the Vicsek model suggest that flocking systems exhibit anomalous fluctuations $\left\langle \delta N^{2}\right\rangle \sim\left\langle N\right\rangle ^{8/5}$ \cite{Ginelli2008, Ginelli2016}. However, the experimental observation of GNF in flocking biological systems has been so far quite scarce and limited to systems showing nematic order \cite{ChateNema}. Also, while the flocking transition has been thoroughly investigated in low density systems \cite{Vicsek95, TonerTu1,TonerTu2, Ginelli2008, Ginelli2016, Gregoire2004}, it is not yet fully understood how it occurs in systems characterised by a large density, such as confluent cellular monolayers. In particular, little is known about how the flocking transition affects the structural, physical properties from the microscopic to the mesoscopic scales. If from a biological standpoint, one is interested in better understanding the microscopic features that control the transition to collective motion, the physical properties emerging from the interplay between jamming \cite{Angelini22032011}, individual active deformation \cite{Zehnder,PhysRevE.92.032729,2016arXiv160701734T}  non-equilibrium clustering effects \cite{MIPS,Levis2014} and the spontaneous symmetry breaking underlying the flocking transition are still not completely understood \cite{Farrel}, especially at the finite (and often relatively small!) length and time scales of biological relevance.

In this work, we investigate experimentally the anomalous number fluctuations and spatial correlations occurring in highly confluent epithelial cell monolayers undergoing a flocking transition to polar order. As a model system, we study doxycycline-inducible, RAB5A-expressing populations of human mammary epithelial MCF-10A cells. RAB5A is a master regulator of the endocytic activity of the cell and it plays an important role in the dissemination of aggressive breast tumours \cite{Frittoli14,Mendoza14,Zhao10}. In control monolayers, cell density increases due to mitotic division, which causes a near complete kinetic arrest akin to a jamming or rigidity transition \cite{Angelini22032011,sadati2014}. We have recently shown that, under these latter conditions \cite{Malinverno2017}, the elevation of RAB5A reawakens the motility of kinetically arrested monolayer, by promoting a flocking transition accompanied by the emergence of large-scale, collective, directed migration patterns \cite{Vicsek201271,Vicsek95}.

Here we study in more detail this flocking transition and its dramatic impact on the structural properties of the monolayer. In agreement with low density theoretical predictions, the onset of an ordered migrating state is found to be accompanied by the emergence of GNF that we also monitor as a low-$q$ divergence in the monolayer static structure factor $S(q)$. We also find that in the reawakened monolayer, the small scale behaviour of $S(q)$ shows that the positional ordering of the cells is less pronounced compared to the near-jammed control, indicating significative fluidisation effects.

We interpret these results through the well-known Toner and Tu hydrodynamic theory of flocking \cite{TonerTu1,TonerTu2}, and by making use of a minimal model of self-propelled soft disks, first introduced in \cite{Szabo, Silke}. Our numerical results show that a simple self-propelled mechanical model with no adhesion forces is able to reproduce qualitatively and even quantitatively the static structure factor and the number fluctuations of both the highly confluent reawakened samples and the near-jammed controls. Furthermore, in agreement with experimentally motivated conjectures on the effects of RAB5A \cite{Malinverno2017}, the transition between these two phases can be triggered, at constant cell density, by decreasing the time scale by whom cells orient their self-propulsion with their local velocity. Finally, our numerical simulations suggest that RAB5A over-expression could also indirectly overcome the feedback mechanisms, such as contact-inhibition of locomotion \cite{Abercrombie:1953jb}, suppressing individual cell motility  in the high-density, near-jammed phase.

\section{Methods}

\subsection{Cell culture and transfection}
MCF-10A cells were a kind gift of J. S. Brugge (Department of Cell Biology, Harvard Medical School, Boston, USA) and were maintained in Dulbecco\textquoteright s Modified Eagle Medium: Nutrient Mixture F-12 (DMEM/F12) medium (Invitrogen) supplemented with $5\%$ horse serum, $0.5$ mg/ml hydrocortisone, $100$ ng/ml cholera toxin, $10$ \textmu g/ml insulin and $20$ ng/ml EGF. The cell line was authenticated by cell fingerprinting and tested for mycoplasma contamination. Cells were grown at $37$ \textdegree C in humidified atmosphere with $5 \% $ CO2 . MCF-10A cells were infected with pSLIK-neo-EV (empty vector control) or pSLIK-neo-RAB5A lentiviruses and selected with the appropriate antibiotic to obtain stable inducible cell lines. Constitutive expression of EGFP-H2B was achieved by retroviral infection of MCF-10A cells with pBABE- puro-EGFP-H2B vector. 

\subsection{Microscopy observation}
\begin{figure*}
\includegraphics[scale=0.5]{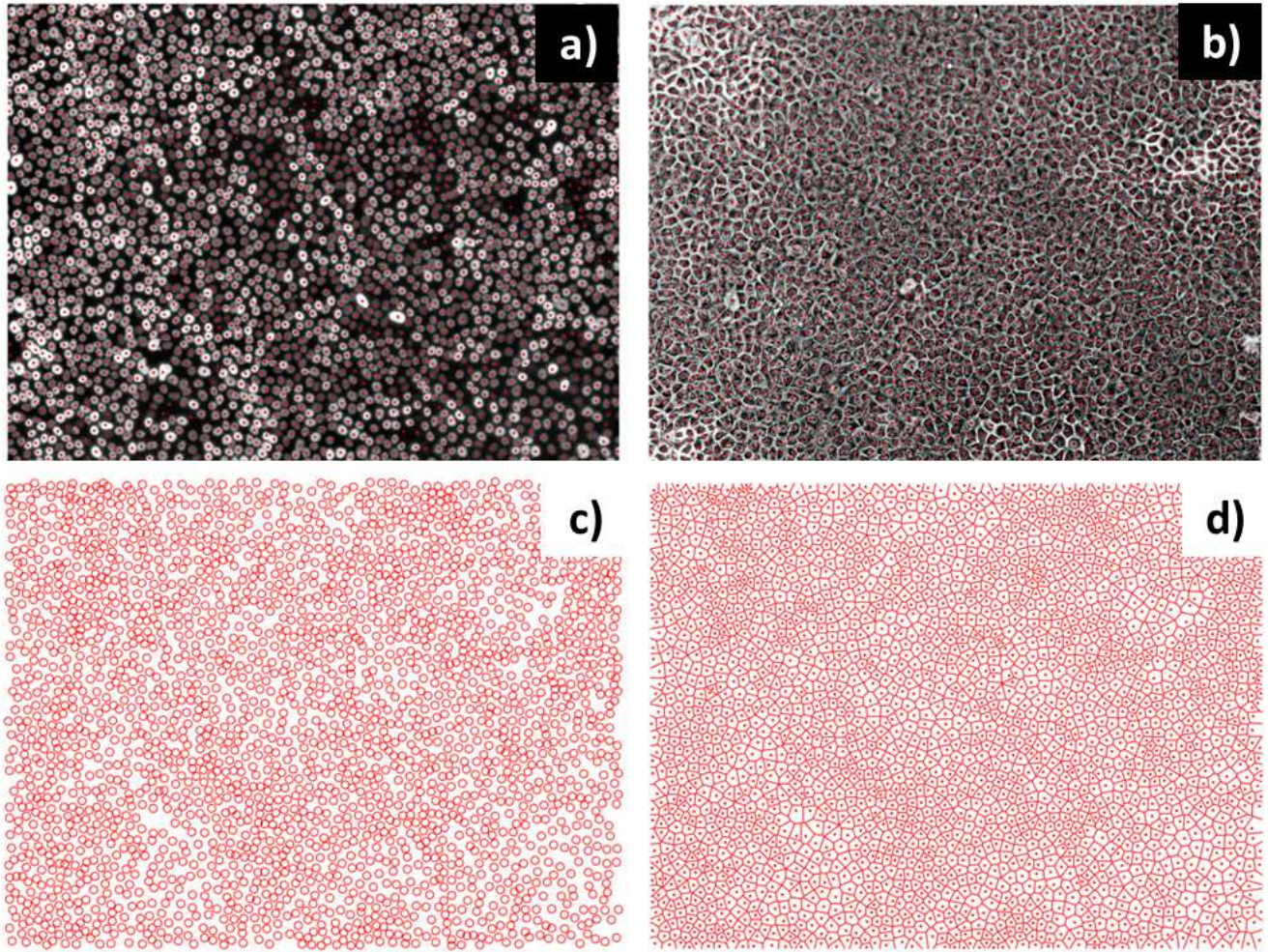}\caption{\label{fig:imaging_modes}a) Representative fluorescence microscopy image of a highly-confluent control epithelial cell monolayer. The horizontal size of the image corresponds to 1.3 mm. Cell nuclei are fluorescently tagged and their positions can be determined with high accuracy (red dots). b) Same field of view of panel a) under phase-contrast imaging. (c) Reconstruction of the nuclei distribution, represented as disk having approximately the same size of the nuclei. (d) Voronoi diagram of the nuclei centres. }
\end{figure*}

Cells were seeded in six-well plates ($1.5\times10^{6}$ cells/well) in complete medium and cultured until a uniform monolayer had formed. RAB5A expression was induced, where indicated, $16$ h before performing the experiment by adding fresh complete media supplemented with $2.5$ \textmu g/ml doxycycline to cells. At the time of recording, fresh media containing EGF was added. An Olympus ScanR inverted microscope with \texttimes 10 objective was used to acquire images with a frame rate of $10$ min over a $24$ h period. For each condition (RAB5A and control) 5 independent fields of view (FOVs) (much smaller than the entire culture plates and far from the boundaries) are captured. Each FOV is imaged both in fluorescence and in phase-contrast (Fig.\ref{fig:imaging_modes}(a)-(b)). The assay was performed using an environmental microscope incubator set to $37$ \textdegree C and 5$\%$ CO2 perfusion. After cell induction, doxycycline was maintained in the media for the total duration of the time-lapse experiment.

\subsection{Image processing, monolayer static structure factor and number fluctuations}
\label{measures}

To quantitatively characterise the structural properties of the cell spatial arrangement, we measure the isotropically-averaged static structure factor $S(q) \equiv \langle S(\boldsymbol{q}) \rangle_{|\boldsymbol{q}|=q}$ of the monolayer, obtained as the orientational average $\langle \cdot \rangle_{|\boldsymbol{q}|=q}$ of the full static structure factor

\begin{equation}
S(\boldsymbol{q})=\frac{1}{N_{0}}\left\langle \sum_{n,m}e^{-j\boldsymbol{q}\cdot\left(\boldsymbol{x}_{n}-\boldsymbol{x}_{m}\right)}\right\rangle. \label{eq:vv}
\end{equation}

Here $N_{0}$ is the total number of cells, $\boldsymbol{x}_{n}$ the position of the $n$-th cell (identified as the centroid of its nucleus) and $\left\langle \cdot\right\rangle$ indicates an average over different experimental realisations and, in the case of a stationary system, over a prescribed time window. Introducing the number density: $\rho(\boldsymbol{x})=\sum_{n}\delta(\boldsymbol{x}-\boldsymbol{x}_{n})$, and the fluctuation $\delta\rho(\boldsymbol{x})=\rho(\boldsymbol{x})-\rho$ around the average density $\rho,$ the static structure factor can be also written as 
\begin{equation}
S(\boldsymbol{q})=\frac{1}{N_{0}}\left\langle |\hat{\delta\rho}(\boldsymbol{q})|^{2}\right\rangle \,,
\label{Sq}
\end{equation}
where $\hat{\bullet}$ indicates the 2D Fourier transform operation. As it can be appreciated in Fig. \ref{fig:imaging_modes}(a), there is a certain degree of variability both in the size and, more importantly, in the fluorescence intensity of the nuclei, which makes a fully automated tracking of all the nuclei rather difficult. To overcome this problem, we adopted an operator-supervised, semi-automatic procedure. Each image is first processed with a standard, centroid-based particle tracking algorithm \cite{Xiao2016157}, whose results are then carefully checked cell by cell and manually corrected with the help of a custom software developed in MATLAB. The data considered in this study consist of image sequences, acquired over a time window of $24$ hours with a frame rate of  six images per hour,  of five independent FOVs for each of the two conditions (control and RAB5A). For each FOV, three images, evenly distributed in the time window comprised 5 and 15 h, are selected and analysed. As can be appreciated also from Fig. \ref{fig:imaging_modes-1-1}, within this time window the macroscopic parameters describing cell motility maintain an almost constant value and the system could be assumed to be in an approximately stationary state. In total, 30 images (with an average of about 2500 cells each) are used to increase the statistical accuracy of our analysis.

Beside evaluating the static structure factor, a very direct way to quantify the degree of spatial homogeneity in the distribution of discrete entities is to consider the fluctuation $\delta N=N-\left\langle N\right\rangle $ in the number $N$ of cells contained in a region of prescribed size around its mean value $\left\langle N\right\rangle $. We measure $\left\langle \delta N^{2}\right\rangle $ as a function of $\left\langle N\right\rangle $ by counting the cells enclosed in non-overlapping square boxes of linear size $a$ variable in the range $[1.3,\,300]$ $\mu m$.

The statistics of number fluctuations is intimately related to the static structure factor. In fact, the number of cells in a given region $A$ (of area $a^2$) of the plane can be written as $N=\int_{A}d^{2}\boldsymbol{x}\rho(\boldsymbol{x})$. Correspondingly, the mean value of the cell number is $\left\langle N\right\rangle =a^2\rho$ and the mean squared value is given by

\begin{equation}
\left\langle \delta N^{2}\right\rangle =\frac{\left\langle N\right\rangle }{a^2}\int d^{2}\boldsymbol{q}|\hat{\chi}_{A}(\boldsymbol{q})|^{2}S(\boldsymbol{q}),\label{dn_S}
\end{equation}

where $|\hat{\chi}_{A}(\boldsymbol{q})|^{2}$ is the Fourier spectrum of the characteristic function of the region $A$. For example, if $A$ is a square of side $a$, we have: $|\hat{\chi}_{A}(\boldsymbol{q})|^{2}=a^{4}\text{sinc}^{2}(\frac{q_{x}a}{2})\text{sinc}^{2}(\frac{q_{y}a}{2})$. 

From Eq. \ref{dn_S} it is clear that the behavior of $\left\langle \delta N^{2}\right\rangle $ for large $\left\langle N\right\rangle $ is determined by the behaviour of $S(q)$ in the $q\rightarrow0$ limit, that is 
\begin{equation}
 \frac{\left\langle \delta N^{2}\right\rangle }{\left\langle N\right\rangle } \sim S(q \to 0),
\label{GDF}
\end{equation}
in the large particles number limit. In fact, if a finite limit exists for $S(q)$ for $q\rightarrow0$, one has normal fluctuations i.e. $\left\langle \delta N^{2}\right\rangle \sim \left\langle N\right\rangle$. This is the case for instance of equilibrium systems, for which $S(q \to 0)=\rho k_{B}T\chi_{T}$, where $k_{B}$ is the Boltzmann constant, $T$ is the absolute temperature and $\chi_{T}$ is the isothermal compressibility \cite{RevModPhys.85.1143}.

On the contrary, if $S(q)$ diverges algebraically for small $q$, $S(q)\sim q^{-\alpha}$ with $\alpha>0$, the scaling of the mean squared amplitude of the number fluctuations is no longer linear in the mean number, and anomalously large number fluctuations are present \cite{RamaswamyEPL}: $\left\langle \delta N^{2}\right\rangle \sim\left\langle N\right\rangle ^{1+\frac{\alpha}{2}}$. As shown by the Toner and Tu hydrodynamic theory \cite{TonerTu1,TonerTu2}, this is the case of active matter systems displaying collective motion. The spontaneously broken rotational invariance of the flocking state is, in fact, subject to large wavelength orientation fluctuations which are easily excited and decay very slowly (these are the celebrated Nambu-Goldstone modes \cite{Nambu,Gold}). Due to the non-equilibrium, self-propelled nature of active matter, density fluctuations are coupled to the orientational ones, so that not only the  velocity fluctuations correlations, but also the density fluctuation ones become long-ranged. Correspondingly, the density static structure factor $S(q)$ shows a divergence for $ q\to 0$. While in the hydrodynamic theory $S({\bf q})$ is characterised by a somehow complicated anisotropic behaviour, it is possible to compute its isotropically averaged small wavenumber behaviour. For polar aligning systems in two spatial dimensions, it is predicted to diverge as $S(q) \sim q^{-\alpha}$ with an exponent $\alpha=6/5$ \cite{Ginelli2016}. This finally gives the anomalous fluctuations $\left\langle \delta N^{2}\right\rangle \sim\left\langle N\right\rangle ^{8/5}$ \cite{Ginelli2016}, as confirmed by numerical simulations of the Vicsek model \cite{Ginelli2008}. In the disordered non-flocking phase, a finite $S(0)$ and normal fluctuations are expected from the theory in the large scale, hydrodynamic limit. However, in active systems with repulsive interactions one expects non-equilibrium clustering effects, which could also lead either to true phase segregation phenomena or to anomalously large fluctuations at some mesoscopic, intermediate scale \cite{MIPS, Levis2014, Silke2}.

\subsection{Quantification of cell motility and orientational order }
Coarse-grained maps of the instantaneous velocity within the monolayer are obtained by performing a Particle Image Velocimetry (PIV) analysis on the phase-contrast image sequences, as described in detail in \cite{Malinverno2017}. For each image, vectorial velocities are obtained over a square grid with spacing of $32$ pixels, corresponding to $20.6$ $\mu m$. We indicate with $\boldsymbol{v}_{j}^{(k)}(t)$ the vectorial velocity measured at time $t$ in the $j$- th grid point of the $k$-th FOV. The overall motility of the cells within the monolayer at a given time $t$ is well captured the mean squared velocity $v_{RMS}(t)$, defined as:

\begin{equation}
v_{RMS}(t)=\sqrt{\left\langle\frac{1}{M}\sum_{j=1}^{M}|\boldsymbol{v}_{j}^{(k)}(t)|^{2}\right\rangle_k}\,,
\label{vrms}
\end{equation}

while the degree of orientational order is quantified by the instantaneous
order parameter $\psi(t)$, calculated as \cite{Vicsek95}:

\begin{equation}
\psi(t)=\left\langle\left|\frac{1}{M}\sum_{j=1}^{M}
\frac{\boldsymbol{v}_{j}^{(k)}(t)}{|\boldsymbol{v}_{j}^{(k)}(t)|}\right|\right\rangle_k\,.
\label{eq:psi}
\end{equation}

In the above expressions $\langle \cdot \rangle_k$ denotes an average over different (and statistically independent) FOVs (five in the present case) and $M$ is the number of grid points per image ($M=1024$ in our case). According to Eq.\ref{eq:psi}, $\psi$ is comprised between $0$ and $1$. If all the cells within each FOV coherently migrate in the same direction one has $\psi\simeq1$, while $\psi\simeq 1/\sqrt{M}$ is expected for the case of a randomly oriented velocity field.

\subsection{Numerical active matter model}
To simulate the epithelial monolayer we use a collisional Vicsek-like model  (CVM), first introduced in \cite{Szabo, Silke}. In this agent-based model, $N_0$ soft disks with position ${\bf r}_i$ and interaction radius $\sigma_i$ $(i=1,\ldots,N_0)$, move with self-propelled speed $v_0$ in the unit direction $\hat{\bf n}(\theta_i)=(\cos \theta_i, \, \sin\theta_i)$ (the particle polarisation) and interact via harmonic repulsive forces, according to the following overdamped dynamics
\begin{equation}
\dot{\bf r}_i = v_0 \hat{\bf n}(\theta_i) + \beta \sum_j^{N_0} {\bf F}_{ij}
\label{CVM1}
\end{equation}
where $\beta$ controls the intensity of harmonic repulsion (akin to a cell membrane tension), and where the repulsive force may be written as
\begin{equation}
{\bf F}_{ij} = \left\{\begin{array}{c r}
0 \;\;\;& \mbox{if}\;\; r_{ij}\equiv|{\bf r}_i - {\bf r}_j| \geq (\sigma_i+\sigma_j)\\
\left[ r_{ij} - (\sigma_i+\sigma_j)\right]\hat{\bf r}_{ij}  \;\;\;& \mbox{if}\;\; r_{ij} < (\sigma_i+\sigma_j)
\end{array}\right.\,,
\end{equation}
with $\hat{\bf r}_{ij}$ being the unit vector from particle $i$ to $j$. The soft disk radii can be polydisperse, and we choose them to be equally distributed in the interval $[1-p/2,\,1+p/2]$, with $p$ a polydispersity parameter. 

In the simplest implementation of this model \cite{Silke2}, the particle polarisation angles $\theta_i$ evolve randomly according to $\dot{\theta}_i=\xi_i$, where $\xi_i$ is a zero-average and uniformly distributed white noise with standard deviation $\eta$,
\begin{equation}
\begin{array}{c}
\langle \xi_i \rangle = 0\\
\langle \xi_i(t) \xi_j(t') \rangle=\eta^2 \delta_{ij}\delta(t-t')\,.
\end{array}
\end{equation}
However, it is possible to obtain a transition towards collective motion by including the tendency of the cell polarization $\hat{\bf n}(\theta_i)$ to align towards its actual velocity ${\bf v_i}\equiv \dot{\bf r}_i = v_i \, (\cos \psi_i,\, \sin\psi_i)$, that is
 \begin{equation}
\dot{\theta}_i = \frac{1}{\tau} (\theta_i - \psi_i) + \xi_i
\label{CVM2}
\end{equation}
where $\tau$ is the realignment timescale \cite{Szabo}. 

A final, relevant control parameter for the CVM is obviously the packing fraction
\begin{equation}
\phi = \rho \pi \langle \sigma_i^2 \rangle \approx \rho \pi \left(1 + \frac{p^2}{12}\right)\,.
\label{phi}
\end{equation}
where we have used the fact that for $N_0 \gg 1$ one has $\sigma_i^2 = (1 + p^2/12)$. We have chosen a length unit such that the mean disk radius $\bar{\sigma}$ is set to one. We also rescale time units to get rid of a second model parameter, setting the force constant $\beta=1$.
In the $\tau \to \infty$ limit, the CVM is a model for non-ordering active brownian particles and is characterised by a 
phase diagram showing a liquid phase with non-equilibrium clustering, phase separation and a completely jammed 
state \cite{Silke2}. For finite $\tau$ (and not too large noise values) the CVM is known to display a transition from a 
disordered gas-like state to a flocking state \cite{Szabo} (i,e, an ordered polar liquid) 
characterised by GNF \cite{Silke} as expected by 
hydrodynamic theory. A completely jammed state, finally, is still present at high packing fraction and low self
propulsion speed \cite{Silke}. In this work, we will not explore this part of the phase diagram, limiting ourselves to regions where the system is in a
gas-like or liquid-like state, both in the presence and in the absence of polar order.

Numerical simulations have been performed by a simple Euler-Maruyama method with time-steppings
$\Delta t= 10^{-2} \sim 2 \cdot 10^{-3}$\footnote{Chosen appropriately according to the magnitude of the velocity and to mean particles
radius ratio. We have also verified that our numerical results are unchanged (in the statistical sense) when a smaller step size is selected.} 
in a square domain with periodic boundary conditions and linear system size $L$.

For simplicity, we measure the instantaneous orientational order $\psi$ making use, in Eq. (\ref{eq:psi}), of the instantaneous
particle velocities ${\bf v}_i(t)$ rather than working with spatially coarse-grained quantities as one is forced to do in 
experimental measures. On the other hand, to better mimic the experimental procedure, 
we find more appropriate to measure the particles mean squared velocity
$v_{RMS}(t)$ measuring the actual displacement over a finite time $t_D$ ($t_D=50$ time units in our case) and 
plugging in Eq. (\ref{vrms}) the discrete velocity $\Delta {\bf v}_i(t) = [{\bf r}_i(t+t_D)-{\bf r}_i(t)]/t_d$.

The static structure factor is measured through Eq. (\ref{Sq}), where the coarse-grained density field $\rho({\bf x})$
is computed over a fine grid (of linear size $\ell \approx 0.5)$ in order to resolve the small scale details. 
The isotropically-averaged structure factor $S(q)$ is then further averaged by sampling the stationary state dynamics
at periodic intervals (every $t_D$ time units). The same stationary dynamics sampling is also used for number fluctuations, which 
are measured following the box counting procedure outlined in section \ref{measures}.
In this work we present results from simulations of the rescaled ($\bar{\sigma}=1$, $\mu=1$) CVM (\ref{CVM1})-(\ref{CVM2}) in a square domain of side $L=192$. To model a confluent tissue, we work with packing fractions $\phi>1$, yielding typical particles numbers $N_0 \gtrsim 1.2 \cdot 10^4$. While this is at least 4 times larger than the number of cells imaged in the experimental FOV, it should be recalled that in the experiments the monolayer is much larger than a single FOV. Thus, our choice minimises boundary effects at the experimental FOV scales while keeping the computational effort largely  manageable. Also, notice that while we aim at a multi-scale description of the epithelial tissue structure, in our numerical simulations we are more interested in the local and mesoscopic scales, rather than in the hydrodynamic behaviour. The latter, which can be accessed at large scales, does only depends on general symmetries and conservation laws. Therefore, it is generically shared by all polar flocking models in the Toner \& Tu phase (\textit{i.e.} the so-called Vicsek universality class). Larger scale simulations ($N_0 \sim 10^6$) will be reported elsewhere \cite{unpublished}. Based on visual inspection of cell samples we also introduce a 20 $\%$ polydispersity in the cell interaction radius ($p=0.4$) and, somehow arbitrarily, we fix the noise standard deviation at $\eta=0.45$. This leaves us with only 3 free parameters for our model: the packing fraction $\phi$, the velocity $v_0$ and the realignment timescale $\tau$.

\section{Results and discussion}

\subsection{Experiments}

RAB5A expression did not alter significantly the migration of individual MCF-10A cells. In Ref.\cite{Malinverno2017} the single-cell behaviour of both RAB5A overexpressing cell and control cells has been characterized in random migration assays in terms of mean velocity, directional persistence and extension of protrusions. For none of these indicators a significant difference was found between control and RAB5A cells. In particular, in both cases, a mean velocity of around 60 µm/h was measured. Thus, RAB5A effects on motility are emergent properties of cell collectives that elicit reawakening of locomotion of kinetically arrested, jammed epithelia. A detailed description of the effects of the over-expression of RAB5 on the kinetics of the confluent MCF-10A monolayer can be found in \cite{Malinverno2017}. As it can be appreciated from Fig. \ref{fig:imaging_modes-1-1}(a), as well as in supplementary movies M1\_PC and M1\_FL, the overall cellular motility in the control monolayer is substantially suppressed and no trace of cell-cell orientational order can be found ($\psi<0.1$,  Fig. \ref{fig:imaging_modes-1-1}(c)). On the contrary, in the same observation window, RAB5A over-expression induces a spectacular reawakening of the cellular motility leading to long range orientational order \cite{Malinverno2017} and a 20-fold increase in the absolute velocity (see Fig. \ref{fig:imaging_modes-1-1}(b)). Moreover, the vectorial velocity field shows a high degree of polar order ($\psi\gtrsim0.9$ in Fig. \ref{fig:imaging_modes-1-1}(d)), indicating a strong cooperativity in the cell migration pattern, typical of a flocking state. The decrease of motility observed in the last part of the experiment can be mainly ascribed to the exhaustion of the Epithelial growth factor (EGF) in the cell culture media.  In fact, EGF is strictly necessary not only for the growth but also for the motility of MCF-10A cells \cite{Deb03,Malinverno2017,Mut01,seton04}

The structural signatures associated with this phenomenology are described in detailed in the following paragraphs.

\begin{figure*}
\includegraphics[scale=0.85]{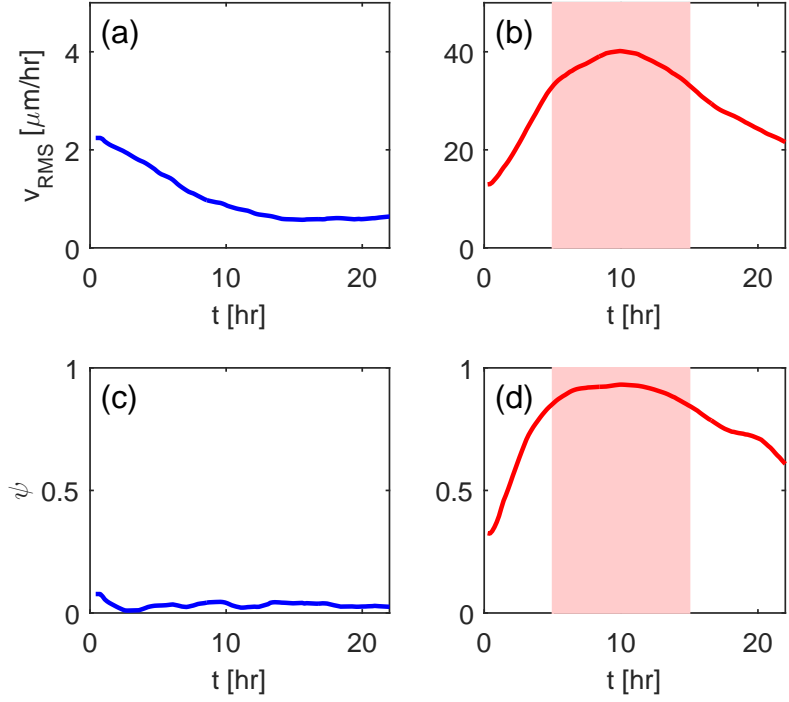}\caption{\label{fig:imaging_modes-1-1}(a) Time evolution of the mean square
velocity for the control monolayer during the experimental observation,
showing a progressive slowing down of cellular motility. (b) Mean
square velocity for RAB5A monolayer. The wide peak in $v_{rms}$ (dashed area) coincides
with a very large value assumed by the corresponding orientational
order parameter $\psi$ (panel (d)), indicating the RAB5A-induced
reawakening of motility coincides with the emergence of a directed
migratory pattern (flocking). The dashed area marks the time window 5-15 h over which stationary averages were taken.
In stark contrast, no orientational
order can be detected in the control monolayer (panel (c)), for which
$\psi\simeq0$.}
\end{figure*}

\subsubsection*{Control cells}

\begin{figure*}[t!]
\centering
\includegraphics[width=0.85\textwidth, clip=true]{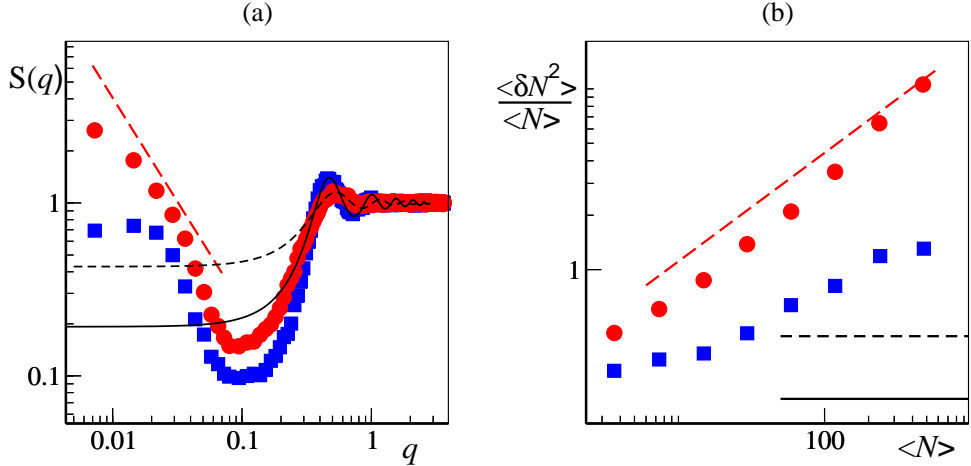}\caption{\label{fig:imaging_modes-1}
(a) Blue squares: static structure factor $S_{0}(q)$ of the control monolayer. The continuous line corresponds to the structure factor of an equilibrium ensemble of randomly scattered hard disks of diameter $d_{0}=11.6$ $\mu m$ and surface fraction $\phi_{0}=0.45$. This model is in fair agreement with the experimental data for large $q$ but does not account for the behaviour of $S_{0}(q)$ in the lower-$q$ regime, indicating the presence of non-equilibrium clusters. Red circles: structure factor $S_{1}(q)$ of the RAB5A monolayer. The dashed line corresponds to the structure factor for a  monodisperse ensemble of hard disks of radius $d_{0}=10.0$ $\mu m$ and surface fraction $\phi=0.25$. The marked divergence exhibited by $S_{1}(q)$ for $q\rightarrow0$ is the signature of the presence of giant density fluctuations that are not compatible with equilibrium models. The dashed red line marks the theoretical prediction flocking systems, $S_{1}(q)\sim q^{-6/5}$. (b) Normalised mean squared number fluctuations for the control (blue squares) and the RAB5A monolayer (red circles), respectively. The dashed red line marks the theoretical prediction (see text). The horizontal lines represent the asymptotic behaviours expected for large $\left\langle N\right\rangle$ by assuming the two hard disks models whose static structure factors are shown (with the same line types) in panel (a).} 
\end{figure*}

The isotropically-averaged static structure factor $S_{0}(q)$ of the control monolayer is shown in Fig. \ref{fig:imaging_modes-1}(a) (blue squares), while a representative snapshot of the cell distribution is reported in Fig. \ref{fig:imaging_modes}. As it can be appreciated from Fig. \ref{fig:imaging_modes-1}(a), in the high-$q$ regime (that is for $q\gtrsim d^{-1},$where $d\simeq12$ $\mu m$ is a typical intercellular distance) the presence of a well defined set of damped oscillations clearly indicates that a partially spatially ordered liquid-like structure exists, strongly reminiscent of that of an assembly of repulsive particles at intermediate density. For comparison, the continuous line in Fig. \ref{fig:imaging_modes-1}(a) represents the static structure factor calculated for an equilibrium ensemble of hard disks (HDs) with diameter $d_{0}=11.6$~$\mu m$ and surface fraction $\phi_{0}=0.45$ \cite{addabedia}. This curve is in very good agreement with the experimental data for large $q$. Interestingly enough, the associated number density $\rho_{0}=\frac{4\phi_{0}}{\pi d_{0}^{2}}=4.3\,10^{-3}$ $cells/\mu m^{2}$ matches almost perfectly the experimentally measured value: $\rho^{0}_{exp}=(4.3\pm0.1)\,10^{-3}$ $cells/\mu m^{2}$. Moreover, the HD diameter $d_{0}$ is fully compatible with the average size $d_{exp}\simeq11\pm1$ $\mu m$ of the nuclei, as obtained from a direct measurement on the fluorescence images. Based solely on these results, it would be tempting to say that the local structure of the monolayer could be explained by assuming purely a steric, repulsive interactions between the nuclei. 

However, for smaller wavenumbers, a strong deviation from the behaviour expected for equilibrium HDs is observed. As $q$ is decreased, $S_{0}(q)$ is characterised by a pronounced dip, indicating the presence of mesoscopic-scale inhomogeneities in the cell distribution, or clustering \cite{Angelini22032011,Levis2014}. It is worth noting that this feature cannot appear in equilibrium systems without invoking attractive interactions and for this reason in active systems, one speaks of {\it non-equilibrium clustering}.

Finally, a plateau seems to set in for $q \to 0$, as predicted by the hydrodynamic theory. Correspondingly, the normalised fluctuations $\left\langle \delta N^{2}\right\rangle/\left\langle N \right\rangle $ (blue squares in Fig.\ref{fig:imaging_modes-1}(b)) show an anomalous increase at intermediate scales and eventually seem to approach a constant value for $\left\langle N\right\rangle \gtrsim200$, signalling the onset of normal fluctuations at large length scales, unfortunately quite close to our experimentally accessible observation limit.  An alternative estimate of the typical cluster size beyond which normal fluctuations sets in can be also obtained by considering the shoulder in $S_{0}(q)$, that occurs for $q_{c}\simeq0.025$ $\mu m^{-1}$. The typical number of cells in a cluster can be estimated as $N_{c}\simeq \rho_{exp}(2\pi q_{c}^{-1})^{2}\simeq250$. This estimate is in agreement with the normalised mean squared number fluctuation of the control shown in Fig. \ref{fig:imaging_modes-1}(b), which saturates to a constant value for $\left\langle N\right\rangle \gtrsim200$. It should be noted that, since the control monolayer is almost completely kinetically arrested, these fluctuations are not very dynamic in nature and appear to be substantially ``frozen'' on the experimental timescales (see also Supplementary Movies M1\_PC and M1\_FL). 

\subsubsection*{RAB5A cells}

\begin{figure*}[t!]
\includegraphics[scale=0.5]{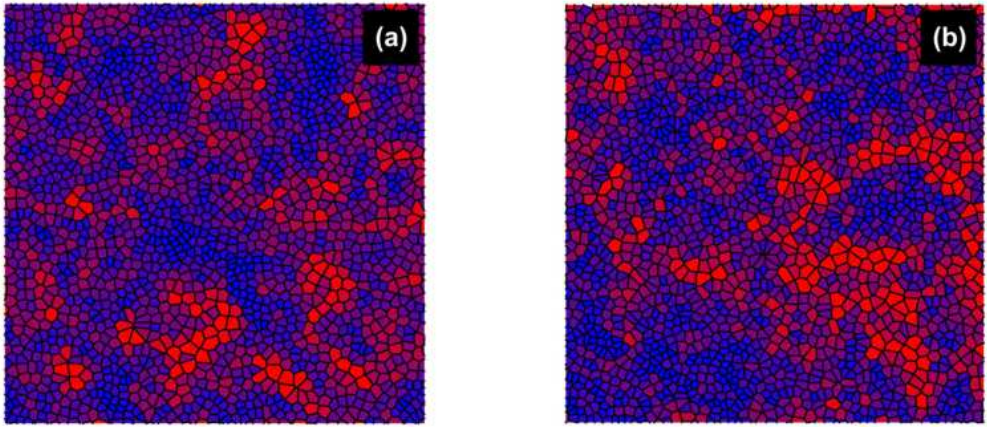}\caption{\label{fig:configurations} (a) Representative snapshot of the CVM in the ordered, high $v_0$ regime (see text) 
and (b) of the RAB5A monolayer. In both panels the Voronoi tessellation generated by the cell centres is shown. Voronoi cells are colour-coded according to their surface area. Each image contains around $N_0 \approx 2000$ cells for a linear size of about $1.0$ mm in real space.}
\end{figure*}

The spatial distribution of RAB5A-overexpressing cells presents a number of peculiar features, as it can be appreciated by inspecting the representative snapshot shown in Fig. \ref{fig:configurations}(b), the supplementary movies M2\_PC amd M2\_FL, and Fig. \ref{fig:imaging_modes-1}(a) (red circles),  where we show the experimentally measured structure factor $S_{1}(q)$. Similarly to the control monolayer, also in this case the high-$q$ portion of the structure factor presents damped oscillations. Compared to controls, the peaks are however less pronounced and slightly shifted to the right, indicating a fluidisation of the monolayer. For reference, the dashed line in Fig. \ref{fig:imaging_modes-1}(a), which is in fair agreement with $S_{1}(q)$ for large $q$, represents the equilibrium structure factor of randomly scattered hard disks, with diameter $d_{1}=10.0$~$\mu m$ and surface fraction $\phi_{1}=0.25$. Compared with the one used for controls, this HD model is characterised by smaller particles, at a lower number density $\rho_{1}=\frac{4\phi_{1}}{\pi d_{1}^{2}}=3.2\,10^{-3}$ $cells/\mu m^{2}$.

It is worth noting that RAB5A over-expression has marginal effects on the rate of cell division of confluent monolayers \cite{Malinverno2017} and, in fact, the measured cell number density $\rho_{exp}^{(1)}=(4.3\pm0.2)\,10^{-3}$ $cells/\mu m^{2}$ is fully compatible with the one of control cells. Moreover, RAB5A has no effect on the size of the nuclei. These findings suggest that RAB5A expression significantly alter the effective cell-cell interaction, namely, by decreasing the minimum distance between neighbouring cells.

At larger scales, a non-equilibrium clustering behaviour analogous to the one exhibited by the control is marked by a dip (albeit less pronounced) in the structure factor. However, in this case a marked divergence is observed for the smallest $q$, which is roughly compatible with the algebraic behaviour predicted by the hydrodynamic theory, $S(q) \sim q^{-6/5}$ in the hydrodynamic regime. 

This behaviour is reflected by the marked CLT violation exhibited by the number fluctuations (red circles in Fig. \ref{fig:imaging_modes-1}(b)), showing, at least up to the larger investigated length scales, a clean algebraically scaling of the normalised number fluctuations $\left\langle \delta N^{2}\right\rangle/\left\langle N\right\rangle \sim \left\langle N\right\rangle^{\gamma}$, with $\gamma=0.66\pm0.04$. This value is quite close to the theoretical value predicted from the hydrodynamic theory for the flocking Toner \& Tu Phase, $\gamma=\sigma/2=3/5$. To the best of our knowledge, this result represent the first clear measure of this anomalous fluctuations in a biological system showing flocking and long ranged polar order.

\subsection{Active matter simulations}

\subsubsection*{Disordered phase}
\begin{figure*}[t!]
\centering
\includegraphics[width=0.7\textwidth, clip=true]{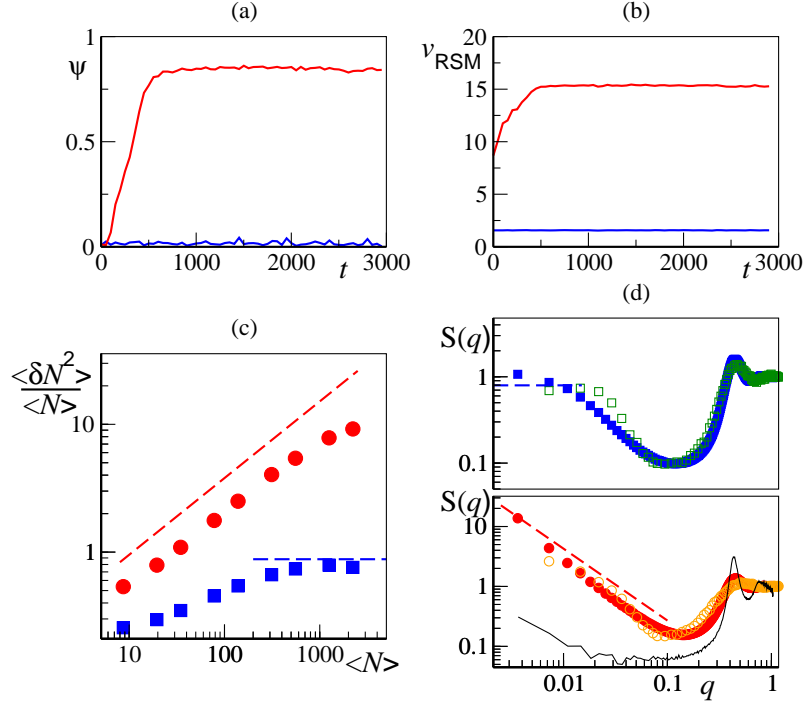}
\caption{\label{Sims}
a) Orientational order parameter $\psi$ vs. time (arbitrary units) for the ordered (red line) and disordered phases
(see text for simulation parameters). 
b) Root mean squared velocity vs. time (both in arbitrary units) for the ordered (red line) and disordered phases.
c) Normalised giant number fluctuations vs. the mean number of particles computed in sub-boxes of increasing 
linear size (see text). Red circles give the ordered phase results, while blue squares are for the 
disordered state. The dashed lines mark the expected hydrodynamic limit theoretical results for, respectively, the
ordered phase (the red line is growing as $\langle N \rangle^{0.6}$) and disordered one (flat blue line).
d) Isotropically averaged static structure factors vs. the wavenumber $q$ (in $\mu m^{-1} units$): 
In the top panel, the numerical $S(q)$, computed in the disordered phase (full blue squares), 
is compared with the control experimental structure factor $S_0(q)$ (open green squares).  
In the bottom panel, the numerical $S(q)$ is computed in the ordered phase (full red circles)
for $v_0=0.2$ and compared with the RAB5A experimental structure factor $S_1(q)$ (open orange circles).  
The black lines shows the numerical structure factor obtained in the ordered regime but at lower
self-propulsion speed ($v_0=0.5$).
The dashed lines mark the expected hydrodynamic limit theoretical results for the two phases (see text).\\
Simulations have been performed with timesteppings $\Delta t=10^{-2}$ (disordered phase) and $\Delta t=2\cdot10^{-3}$ 
(ordered phase). Static structure factors and GNF have been computed averaging over $1000$ different configurations 
of the stationary dynamics (consecutive configurations are separated by $t_D=50$ time units).}
\end{figure*}

For reasons that will be clear in the following, we set the surface fraction to the value $\phi=1.2$ and proceed to optimise the single cell velocity $v_0$ and the orientation time $\tau$ in order to reproduce the control cells experimental results. Good agreement is found for instance for $v_0=0.5$ and $\tau=5.0$. For these values of the parameters, the systems lies in the disordered phase $\psi \approx 2 \cdot 10^{-2}$ (the blue line in Fig. \ref{Sims}(a). Correspondingly, we show in Fig. \ref{Sims}(c) that the normalised number fluctuations (blue squares) converge to a constant plateau for sizes larger than the non-equilibrium clustering size, which indicates normal number fluctuations in agreement with the CLT. 

A qualitative agreement can be found in the disordered, non completely jammed phase for a wide range of model parameters (see also \cite{Levis2014}, where the static structure factor for an ensemble of non-equilibrium self-propelled hard disks is discussed). By contrast, a more quantitative agreement with the experimental static structure factor can be achieved by introducing a physical length scale. We do so by fitting the location of main high $q$ peak of the numerically computed structure factor with the one of the experimental control $S_0(q)$. This provides us with a length scale of one simulation unit being approximately equal to $8.9 \mu m$. In Fig. \ref{Sims}(d) (upper panel) the similarity between the disordered numerical static structure factor (full blue squares) and its experimental counterpart $S_0(q)$ (green open squares) can be fully appreciated over the full multiscale range. We even achieve an almost quantitative agreement ranging from the local, interaction range, structure to the non-equilibrium clustering mesoscopic range and up to the hydrodynamic range in which saturation to a constant is observed.

Moreover, once we express the mean disk interaction radius $\bar{\sigma}$ in the experimental length scale, Eq. (\ref{phi}) gives us a density $\rho \approx 4.7 \cdot 10^{-3}$ $cells/\mu m^{2}$, in reasonable agreement with the experimentally observed one, thus justifyig our initial assumption $\phi=1.2$. A typical sample of the CVM dynamics in the disordered phase can be found in supplementary movie M3.

\subsubsection*{Ordered phase}
For sufficiently low noise amplitudes, it is known that the CVM presents a transition to a polar flocking state as the density is increased or the particle polarity realignment timescale $\tau$ is decreased. As done in \cite{Malinverno2017}, we also chose to keep the noise amplitude and density constant ($\eta=0.45$ and $\phi=1.2$, respectively) and to change the re-orientation time. Keeping all the other parameters unchanged, we verify that the CVM undergoes a transition to collective motion for $\tau \lesssim 2$. We place ourself at a much lower value, namely $\tau=0.45$, also to avoid the spontaneous segregation instability that is known to take place at the onset of collective motion, at least in low density systems \cite{Ginelli2016}. 

Here the dynamics shows clear signs of collective motion, with an order parameter $\psi \approx 0.94$. However, the system shows spatial configurations characterised by a higher degree of positional order compared to the ones observed in the disordered phase. Correspondingly, the structure factor $S(q)$ shown in Fig. \ref{Sims}(d) (bottom panel, black line) is characterised by strong damped oscillations in the high $q$ region. It is easy to understand this phenomenon: when the motion is disordered, the particles often push one against each other, resulting in a lower positional order. In the ordered regime, on the other hand, fluctuations around the mean velocity are smaller resulting in a higher positional order. This is actually the opposite of the fluidisation effects observed in the RAB5A experiments. 

To reproduce the experimental data, we need to decrease the positional order without lowering drastically the orientational one. To this aim, it is necessary to increase the self-propulsion velocity $v_0$. We do so keeping the noise amplitude constant, $\eta=0.45$, and by finding an optimal parameter range for $v_0 \approx 2.0$. The equilibration kinetics and the static structure factor for these parameters are presented in Fig. \ref{Sims}(d), while a typical configuration is shown in Fig. \ref{fig:configurations}(a). Also, a movie of the dynamics can be found as supplementary movie M4.  

Starting from a fully disordered initial condition, the systems quickly aligns, as the order parameter reach its stationary value $\psi \approx 0.84$ (Fig. \ref{Sims}(a), red line). Correspondingly, the root mean squared velocity $v_{rms}$ grows tenfold compared to the one measured in the disordered phase (Fig.\ref{Sims}(b)), in good qualitative agreement with what observed comparing the root mean squared velocities of the RAB5A and the control experimental monolayers.

Due to the increased particle velocities (and the slightly decreased orientational order), positional ordered is significantly lowered, resulting in a notably less sharp peak structure in the high $q$ structure factor, as shown in Fig. \ref{Sims}(d) (bottom panel, full red circles). In fact, one can verify that the main peak is now lower than the numerical disordered data shown in the upper panel.

Overall, our numerical results suggests that in order to achieve both fluidisation and a transition to the flocking state, cells need to i) decrease their polarity reorientation timescale and ii) increase their self-propelled velocity. In our model, indeed, this is achieved by parameter changes of roughly one order of magnitude!

Good qualitative and quantitative agreement with the RAB5A monolayer structure is also observed at smaller $q$, with the numerical $S(q)$ well reproducing the non-equilibrium clustering signatures at the mesoscale as well as the hydrodynamic scale divergence of the RAB5A structure factor (open orange circles in Fig. \ref{Sims}(d)). 

The normalised number fluctuations are finally shown in Fig. \ref{Sims}(c) (red circles). Since we are able to access the hydrodynamic divergence at the relatively small scales of our present simulations (larger scale simulation will appear somewhere else \cite{unpublished}), we are also able to observe asymptotic density fluctuations compatible with theoretical predictions \cite{Ginelli2016}. However, this is not the case for generic parameter values in the flocking phase, and we have verified that for different choices the hydrodynamic range gets pushed at larger length scales, above the one probed by the numerical simulations here reported.

In the above discussion the transition from a disordered state, representative of the control monolayer, to a polarly ordered, fluidised state is obtained by simultaneously decreasing the alignment time $\tau$ and increasing the self-propulsion speed $v_0$, at constant membrane tension $\beta=1$.  Simple rescaling arguments show that changing $\beta$ to values different from unit amounts to the following parameter shifts: $v_0 \to v_0 / \beta$, $\tau \to \beta \tau$ and $\eta \to \eta / \sqrt{\beta}$. This suggests that the same transition can be also triggered by acting on different control parameters, for example by simultaneously decreasing $\tau$ and $\beta$, at constant the self-propulsion velocity. This route is similar to the one originally proposed in Ref. \cite{Malinverno2017} in the context of a self-propelled Voronoi model \cite{PhysRevX.6.021011} with alignment interactions \cite{giavazzipaoluzzi}, where the effect of RAB5A was interpreted in terms of faster cell-cell alignment (lower $\tau$) combined with increased cell-cell adhesion. Beyond this qualitative correspondence, a detailed understanding of the differences (at small and intermediate scales) between the predictions of particle-based models, like the one used in this work, and vertex-like models \cite{Bi2015,PhysRevX.6.021011,Malinverno2017,giavazzipaoluzzi} is still lacking, and it is well beyond the scope of the present work.

\section{Conclusions}

We presented here a study of the structure of a confluent epithelial monolayer switching from an orientationally disordered state, close to kinetic arrest, to a fluidised, polarly ordered state showing directed collective migration. Strikingly, the effects of activity manifest themselves
over a wide range of scales: at large scales, the flocking monolayer shows clear signs of anomalous number fluctuations in agreement with the prediction
of the hydrodynamic theory of flocking. This is the first time these predictions have been unambiguously confirmed in biological active matter showing long ranged polar order. On the local scale, on the other hand, the reawakened cellular motility induces fluidisation effects with a clear structural signature. These findings are at variance with respect to previous studies where the transition from a fluid-like state to an arrested state (jamming) in a confluent epithelial monolayer was primarily described in terms of an increasing length scale associated with the heterogenous dynamics, with marginal or no effects on the static structure \cite{Angelini22032011}.  

To gain further insights on the nature of this transition to fluidised collective motion, we studied numerically a simple mechanical model of soft self propelled disks, first introduced in \cite{Szabo, Silke}, kept at high packing fraction. This simple model, lacking adhesion forces, is able to reproduce qualitatively and even quantitatively the monolayer structure of  both the control and the reawakened RAB5A on a wide range of scales, ranging from the single cell scale to the hydrodynamic collective range. 

While in the context of the model, the transition at constant cell density is mainly controlled by the efficiency of the polarity alignment mechanism, the concurring experimental fluidisation is reproduced by a large increase of the particles self propelling speed. Experimentally, we showed that RAB5A expression does not alter the migration speed of sparsely seeded cells, which is comparable, in absolute terms, to the one characterizing the collective migration of RAB5A cells in a confluent monolayer. Altogether, these results suggest that another effect of the RAB5A induced transition to flocking could be to reduce the mechanically induced feedbacks that tend to suppress cellular motility in the disordered state, such the ones driving the control monolayer close to a kinetically arrested state \cite{Abercrombie:1953jb} This intriguing hypothesis will be tested in future experiments, aimed at dissecting the detailed biochemical and biomechanical mechanisms connecting the overexpression of RAB5A with the emergence of the reported collective behaviour.

In the future, it would be also very interesting to investigate the relationship between the structural effects observed here and the monolayer dynamics at the mesoscopic, biologically relevant scale, such as for instance the effect of the spatial structure on the relaxation of density fluctuations or the coupling between high density elastic modes and velocity fluctuations, or the role played the intracellular adhesion in stabilizing the local structure.

\ack{}{}

We thank S. Henkes for useful discussions. FGia and RC acknowledge funding from the Italian Ministry of University and Scientific Research (MIUR) under the program \textit{Futuro in Ricerca} - Project ANISOFT (RBFR125H0M) and from Regione Lombardia and CARIPLO foundation under the joint action \textit{Avviso congiunto per l'incremento dell'attrattivit\'a del sistema ricerca lombardo e della competitivit\'a dei ricercatori candidati su strumenti ERC}  - Project Light4Life. CM, SC and GS acknowledge funding from Associazione Italiana per la Ricerca sul Cancro (AIRC 10168 and 18621), MIUR, the Italian Ministry of Health, Ricerca Finalizzata (RF0235844), Worldwide Cancer Research (AICR-14-0335), and the European Research Council (Advanced-ERC-268836). CM was also supported by Fondazione Umberto Veronesi and SC by an AIRC fellowship. FGin acknowledges support from the Marie Curie Career Integration Grant (CIG) PCIG13-GA-2013-618399, and wish to thank the University of Milan and LibrOsteria for their hospitality while this work was underway.

\section*{\textemdash \textemdash \textemdash \textemdash \textemdash \textendash{}}

\bibliographystyle{iopart-num}
\addcontentsline{toc}{section}{\refname}
%\bibliography{biblio_xx}

\providecommand{\newblock}{}

%\begin{thebibliography}{24}
% 
%\bibitem{Gabrielli}
%A. Gabrielli, F. Sylos Labini, M. Joyce, L. Pietronero,  
%{\it Statistical Physics for Cosmic Structures} (Springer Berlin, 2005).
%
%\bibitem{unpublished}
%(...) unpublished.
%
%\end{thebibliography}

\end{document}